\documentclass[review]{elsarticle}

\usepackage{lineno,hyperref}
\usepackage{gensymb}
\usepackage{amssymb}

\modulolinenumbers[5]

\journal{arXiv.org}

\begin{document}

\begin{frontmatter}

\title{Impact of the crystal electric field on magnetocaloric properties of CsGd(MoO$_4$)$_2$}

\author[mymainaddress]{V. Tk\'{a}\v{c}\corref{mycorrespondingauthor}}
\ead{tkac.vladimir@upjs.sk}

\author[mymainaddress]{A. Orend\'{a}\v{c}ov\'{a}}

\author[mymainaddress]{R. Tarasenko}

\author[mymainaddress]{M. Orend\'{a}\v{c}}

\author[mymainaddress]{A. Feher}

\cortext[mycorrespondingauthor]{Corresponding author}

\address[mymainaddress]{Institute of Physics, P. J. \v{S}af\'{a}rik University, Park Angelinum 9, 040 01 Ko\v{s}ice, Slovak Republic}

\begin{abstract}
Magnetocaloric effect (MCE) was investigated in the single crystal of CsGd(MoO$_4$)$_2$ in the temperature range from 2 to 30 K and fields up to 5 T applied along the easy and hard magnetic axis. The analysis of specific heat and magnetization provided the refinement of crystal electric field (CEF) parameters supporting the dominance of uniaxial symmetry. The knowledge of CEF energy levels enabled the extrapolation of MCE parameters outside the experimental region. Consequently, maximum values of the isothermal entropy change, $-\Delta S_{\rm M}$,  in magnetic fields up to 5 T are expected to occur at temperatures between 1 and 2 K. While $-\Delta S_{\rm M}$  achieves 19.2 J/kgK already for the field 1 T, for the field change 7 T, maximal $-\Delta S_{\rm M}$ $\approx$ 26.8 J/kgK with a refrigerant capacity of 215 J/kg is expected. The absence of thermal hysteresis and the losses due to eddy currents as well as good chemical stability makes the compound CsGd(MoO$_4$)$_2$ attractive for magnetic refrigeration at low temperatures. The possibilities of further enhancement of MCE parameters are discussed. 

\end{abstract}


\end{frontmatter}


\section{Introduction}

Energy efficient and environmentally friendly technologies receive much attention to solve the energy crisis accompanied with a serious problem of a global warming phenomenon. In this context, refrigeration based on magnetocaloric effect (MCE) has attracted much research interest because of its higher energy efficiency and the use of environmentally friendly materials \cite{Tishin2003,Tegus2002}. MCE resembles processes that occur in gas in response to the changing pressure. Isothermal magnetizing of a magnetic refrigerant reduces magnetic part of the entropy, $\Delta S_{\rm M}$, and corresponds to the isothermal compression of gas. Adiabatic demagnetizing (corresponding to adiabatic expansion of gas) is accompanied with the adiabatic temperature change, $\Delta T_{\rm ad}$. Both thermodynamic quantities describe the measure of MCE. The discovery of a giant MCE in Gd$_5$(Si$_2$Ge$_2$) near a room temperature \cite{Pecharsky1997}, with $- \Delta S_{\rm M}$ $\approx$ 19 J/kg K, for the field change 0 - 5 T triggered flurry of research activities in a wide area of physics, chemistry and material science. 

During last two decades, giant MCE associated with the occurrence of magnetic phase transition of the first or second order has been reported for a broad variety of materials \cite{Pecharsky2002,Wu2015,Aksoy2008,Jayaraman2011,Min2014,Wang2016}. Besides the systems with magnetic phase transition, more options are available for magnetic cryo-cooling. At low temperatures, the vibrational specific heat decreases, consequently, also paramagnets \cite{Sedlakova2009,Chen2013,Chen2014,Guo2013} and superparamagnets \cite{Gschneidner2005} under special conditions which enhance MCE, can be employed. Superparamagnetic systems composed of 3d-4f intermetallic nanoparticles represent nanomagnets with large cryogenic MCE \cite{Zelenakova2016}. However, their application potential is limited due to large content of MCE inactive matrix and in case of encapsulated nanoparticles, their shell has low thermal conductivity \cite{Gschneidner2005}. 
Single molecule magnets or molecular nanomagnets (MNs) represent another class of nanoscopic systems. MNs such as Mn$_{12}$, Fe$_8$ and Mn$_{32}$ containing high-spin molecules (magnetic clusters) with vanishing magnetic anisotropy are particularly favorable due to large magnetic entropy content and easy polarization of net molecular (cluster) spins in magnetic field of low or moderate strength \cite{Evangelisti2009}. 

Currently, much attention is devoted to other isotropic systems, namely Gd-based single-ion magnets (SIMs) \cite{Chen2013,Chen2014,Guo2013}. Unlike aforementioned molecular nanomagnets, SIMs contain paramagnetic ions well separated from others by diamagnetic ligands, thus suppressing the existence of magnetic transitions to very low temperatures. Contrary to conventional bulk refrigerants with magnetic transitions, in molecular nanomagnets, the entropy changes take place in the nanoscale range, thus micron and submicron-sized devices can be fabricated for exploiting the functionality of the molecular coolers \cite{bartolome2013molecular}. Design of effective cryo-refrigerant requires optimization of diverse parameters as magnetic anisotropy, spin value, type and strength of magnetic interactions and relative amount of non-magnetic elements in the structure \cite{Wang2014}. In this respect, molecular-based Gd(III) compounds are usually recognized as good candidates for magnetic refrigeration at low temperatures \cite{Sibille2014}, because Gd(III) ion has zero orbital angular momentum and provides the largest entropy per single ion associated with the highest spin value 7/2. 

Maximal usability of magnetic entropy for the magnetocaloric application is usually reduced by the effects of crystal electric field (CEF) splitting and magnetic correlations characterized by parameters $\Delta$ and $J$. At  relatively high temperatures ($k_{\rm B}$$T$ $\gg$ $\Delta$, $J$) all magnetic degrees of freedom can be used for magnetic cooling. While increasing of density of magnetic atoms enlarges the MCE, it can also reinforce magnetic correlations which lower the magnetic entropy and induce the magnetic ordered state. The solution of this dilemma can be found in the competing effect of magnetic anisotropies from different sources.  As was shown in ref. \cite{Palacios2014}, CEF induced magnetic anisotropy can support the exchange anisotropy which results in the enhancement of the transition temperature $T_{\rm C}$ to the magnetic ordered state and lowering MCE. On the other hand, the competition of the aforementioned anisotropies can suppress the effect of magnetic correlations and $T_{\rm C}$, which  creates appropriate conditions for effective usability of magnetic dense material with maximal magnetic entropy for magnetic cooling.

In this paper we present the study of the magnetocaloric effect in CsGd(MoO$_4$)$_2$. The compound together with other materials from the series of rare-earth double molybdates MR(MoO$_4$)$_2$ (M$^+$ is an alkali-metal ion and R$^{3+}$ is a rare-earth ion) belongs to laser active materials \cite{Shi2014,Zhao2015,Devakumar2017}. Besides their potential use in the optoelectronic devices at room-temperature applications, the compounds possess interesting magnetic properties at low temperatures \cite{Tkac2013,Tkac2014}. Previous low-temperature studies \cite{Stefanyi1988,Feher1988,Zaboj1992,Anders1995,Tibenska2010,Tkac2013} considered this material as a quasi-one dimensional dipolar magnet with a magnetic phase transition to the ordered state at $T_{\rm C}$ = 0.45 K. The effective strength of dipolar interactions was estimated in the form of the effective intra-chain coupling parameter $J/k_{\rm B}$ $\approx$ 0.6 K and interchain coupling $J'$ $<$ 0.03 J \cite{Feher1988}. 

Assuming only dipolar interactions between Gd$^{3+}$ ions, theoretical calculations ~\cite{Anders1995} suggest the presence of the magnetic anisotropy with the easy axis parallel to the crystallographic $c$ axis. On the other hand, the symmetry of the crystal field should induce the anisotropy characterized by the easy axis coinciding with the crystallographic $a$ axis ~\cite{Tkac2014}. Thus, considering the competition between the aforementioned anisotropies,  the weakness of magnetic correlations and CEF splitting as well as the high spin value of Gd$^{3+}$ ion, the maximal usability of magnetic entropy for magnetic cooling can be expected in CsGd(MoO$_4$)$_2$.

The present study is focused at the response of the material to the application of magnetic field investigated in the paramagnetic phase. The analysis of the magnetization and specific heat allowed us to describe basic MCE properties in the studied temperature region. What is more, the refinement of the crystal-field parameters enabled to extrapolate the MCE predictions to a wide parametric region of temperatures and fields, which was not achieved in the current experiment. Our study revealed that besides the aforementioned room-temperature optical applications, this material can be also used as an efficient cryo-refrigerant at helium temperatures already in relatively low magnetic fields $\sim$ 1 - 2 T. 

\section{Crystal structure and experimental details}

CsGd(MoO$_4$)$_2$ is a transparent and soft material which was prepared by a flux method ~\cite{Vinokurov1972} at the Institute of Low Temperature Physics and Engineering in Kharkov. The starting materials, Gd$_2$O$_3$, MoO$_3$ and Cs$_2$O(Cs$_2$CO$_3$) in powder form were weighed according to the stoichiometric ratio and mixed homogeneously. The solid-state synthesis was realized at 700 $^{\circ}$C. The system crystallizes in the $Pccm$ ($D^3_{2 \rm h}$) space group with 2 formula units in the unit cell with the parameters $a$ = 5.07 \r{A}, $b$ = 9.25 \r{A} and $c$ = 8.05 \r{A} \cite{Vinokurov1972,database}. The coordination sphere of Gd$^{3+}$ ion consists of eight oxygen atoms forming a slightly distorted square antiprism (Fig.~\ref{Fig:Models}a). The shortest distance between Gd$^{3+}$ ions is along the c axis, and equals $c$/2. The Gd chains alternate with the chains of [MoO$_4$]$^{2-}$ groups in the $ac$ plane which are separated by Cs$^+$ ions. Thus, only a weak electrostatic coupling exists between the $ac$ planes. 

Isothermal magnetization curves have been investigated in the temperature range from 2 to 30 K in magnetic fields up to 5 T in a commercial Quantum Design SQUID magnetometer. The measurements of the temperature dependence of the specific heat have been performed in the temperature range between 2 and 100 K in the zero magnetic field using commercial Quantum Design PPMS device. Two single crystals in the shape of a thin plate with the mass about 20 mg and 5 mg were used for the magnetization and specific heat measurements, respectively.

Several single crystals with approximate dimensions $a$'$\times$$b$'$\times$$c$' = 2$\times$1$\times$4 mm$^3$ were used for a room-temperature x-ray study to determine the orientation of the crystallographic axes $a$, $b$ and $c$ with respect to the crystal edges $a$', $b$' and $c$'. The longest crystal edge $c$' was identified with the crystallographic $c$ axis, the shortest crystal edge $b$' is parallel to the $b$ axis and the edge $a$' is parallel to the $a$ axis. Crystals grow in the form of thin plates parallel to $ac$ layers with a pronounced lamination within the [010] cleavage plane as a result of very weak coupling between the layers along the $b$ axis.

\section{Results and discussion} 

Gd$^{3+}$ ion has essentially isotropic ground state $^{8}$S$_{7/2}$ ($L$ = 0, $S$ = 7/2, $g_{\rm J} = 2$), which means that no magnetic anisotropy is expected. Consequently, when neglecting dipolar interactions, the magnetocaloric response of the system should depend only on the number of spin degrees of freedom and the density of Gd$^{3+}$ ions. However, a strong spin-orbit coupling of the 4f electrons mixes the ground multiplet $^{8}$S$_{7/2}$ with the first excited state $^{6}$P$_{7/2}$. Consequently, the ground-state contains an admixture of nonzero $L$ states resulting in the crystal-field splitting of the order of 1 K. As will be shown below, even such subtleties can affect magnetization, and corresponding MCE, therefore a special attention was devoted to the crystal-field symmetry in CsGd(MoO$_4$)$_2$.

\subsection{Crystal-field symmetry}

Considering only crystal electric field and magnetic field effects at sufficiently high temperatures, the magnetic behavior of CsGd(MoO$_4$)$_2$ in the paramagnetic phase can be described using Hamiltonian

\begin{eqnarray}
H = H_{CEF} + g_J\mu_BSB,
\label{eq:one}
\end{eqnarray} where $\mu_{\rm B}$ and $S$ stands for the Bohr magneton and spin, respectively. Crystal-field effects are included in the term $H_{\rm CEF}$=$\sum^{2}_{q=0}B_2^q O_2^q+\sum^{4}_{q=0}B_4^q O_4^q +\sum^{6}_{q=0}B_6^q O_6^q$, expressed in Abragam and Bleaney notation \cite{Abragam1970}. $O_k^q$ represents equivalent operators and $B_k^q$ are crystal-field parameters. Previous reports \cite{Tkac2013,Feher1988} estimated crystal-field splitting of the ground state by considering the axial symmetry of the crystal field characterized by the parameter $B_2^0$ = -0.089 K. Corresponding easy axis is parallel to the crystallographic axis $a$ \cite{Tkac2014}.

\begin{figure}
\includegraphics{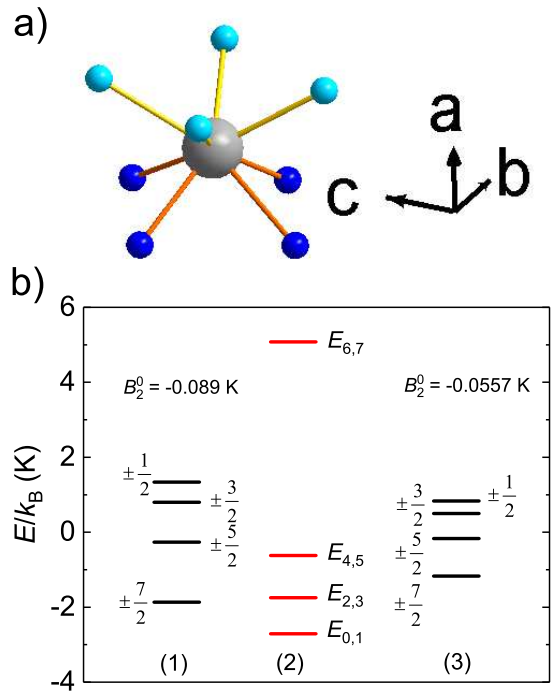} \\
\caption{\label{Fig:Models} (a) The coordination sphere of the Gd$^{\rm 3+}$ ion (gray sphere) with eight oxygen atoms forming a slightly distorted square antiprism with the upper base (light blue spheres) and bottom base (dark blue spheres). (b) Splitting of the ground multiplet $^{8}S_{7/2}$ for different crystal-field symmetries. Energy schemes (1) and (3) correspond to the crystal field with uniaxial symmetry characterized by the parameter $B_2^0$. Energy scheme (2) corresponds to the local symmetry D$_{\rm 2}$ with CEF parameters given in Table.~\ref{tab:table1}}
\end{figure}

\begin{figure}
\includegraphics{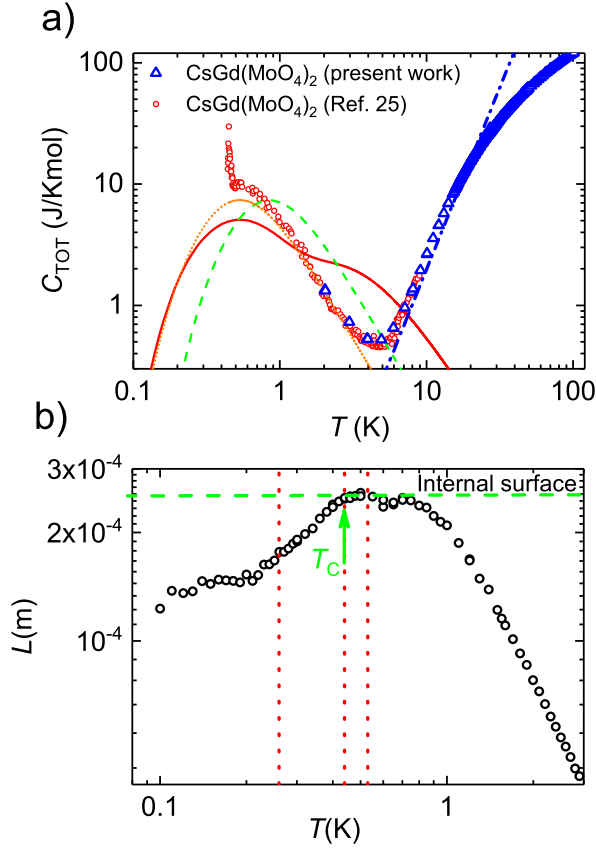} \\
\caption{\label{Fig:HC} (Color online) (a) Temperature dependence of the total specific heat of CsGd(MoO$_4$)$_2$ involving data of present work (triangles) and data taken from Ref. \cite{Stefanyi1988} (circles).  The solid line represents the prediction for the CEF model with parameters from Table 1. Dashed and dotted lines correspond to the CEF model with the axial symmetry for $B_2^0$ = -0.089 K and $B_2^0$ = -0.0557 K, respectively. The dashed-dotted line represents the lattice contribution 1.98$\times$10$^{-3}$$T^3$ \cite{Stefanyi1988}. (b)  Temperature dependence of the phonon mean free path of CsGd(MoO4)2 in zero magnetic field \cite{Tkac2013}. The dashed lines denote hypothetic temperatures corresponding to dominant phonons with the CEF energies for $B_2^0$ = -0.0557 K.}
\end{figure}

To verify the value of the parameter $B_2^0$, the corresponding energy level scheme in Fig.~\ref{Fig:Models}b has been used for the calculation of the specific heat in zero magnetic field. The obtained curve was compared with experimental specific heat depicted in Fig.~\ref{Fig:HC}a. The experimental data set comprises of previous low-temperature data below 2 K \cite{Stefanyi1988} completed with the current measurements performed above 2 K. It is clear that the description of the data is not sufficient, pointing at the overestimation of the CEF influence. Thus, the effect of magnetic correlations and/or other low-symmetry terms in $H_{\rm CEF}$ should be included. 

\begin{figure}
\includegraphics{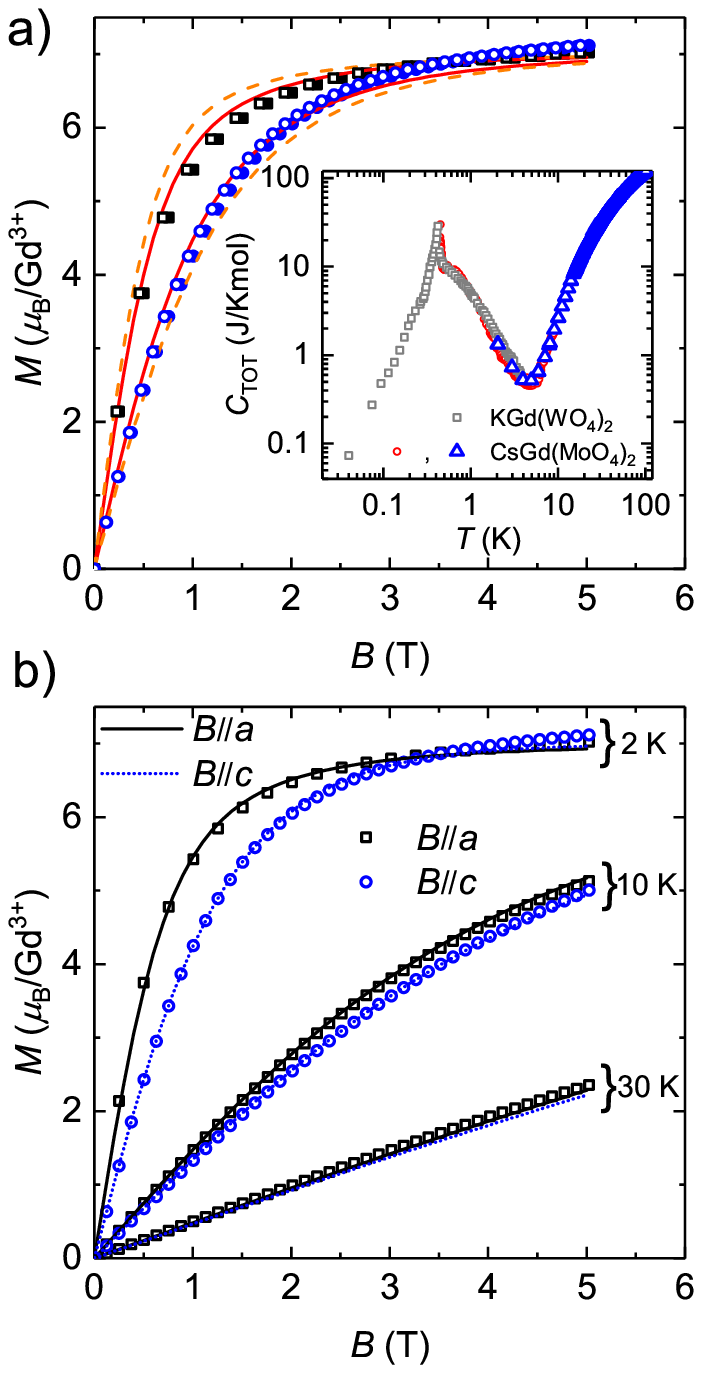} \\
\caption{\label{Fig:Magnetisations} (Color online) (a) Isothermal magnetization curves of CsGd(MoO$_4$)$_2$ for $B$$\parallel$$a$ (full and open squares) and for $B$$\parallel$$c$ (full and open circles) measured at $T$ = 2 K. The open symbols represent data corrected for demagnetization effect with the average demagnetization factor $N$ = 0.2. The lines correspond to theoretical predictions for the CEF model with the axial symmetry for $B_2^0$ = -0.089 K (dashed) and  $B_2^0$ = -0.0557 K (solid).  Inset: Temperature dependence of total specific heat. Open squares represent data of KGd(WO$_4$)$_2$ taken from Ref. \cite{Borowiec2013}, open circles and open triangles have the same meaning as in Fig.~\ref{Fig:HC}a. (b). Isothermal magnetization curves of CsGd(MoO$_4$)$_2$ for $B$$\parallel$$a$ (open squares) and for $B$$\parallel$$c$ (open circles) shown at selected temperatures $T$ = 2, 10, 30 K. The lines represent the simultaneous fits with the CEF parameters from Table.~\ref{tab:table1}.}
\end{figure}

To exclude the effect of magnetic correlations which become significant below 1 K, the isothermal magnetization measurements were performed in the paramagnetic phase. The lowest temperature, $T$ = 2 K, corresponds to 4$T_{\rm C}$, which is sufficiently high for neglecting the effect of the magnetic correlations. The magnetization curves obtained in the magnetic field applied along the crystallographic $a$ and $c$ axes are shown in Fig.~\ref{Fig:Magnetisations}.

In accord with the aforementioned CEF induced magnetic anisotropy, the magnetization reflects the presence of a weak anisotropy with the easy axis parallel to the $a$ axis. No hysteresis effects were observed in the experimental data. The magnetization data were compared with the theoretical predictions for $B_2^0$ = -0.089 K (Fig.~\ref{Fig:Magnetisations}a). Alike specific heat, the deviations between the magnetization data and the predictions clearly indicate that the actual anisotropy in CsGd(MoO$_4$)$_2$ is weaker than expected. 

Considering the local symmetry D$_2$ of Gd$^{3+}$ ion determined at room temperature, we realized simultaneous fitting of magnetization curves at various temperatures and for both orientations of magnetic field ($B$$\parallel$$a$ and $B$$\parallel$$c$) using 9 CEF parameters and much better agreement with experimental data was achieved (Fig.~\ref{Fig:Magnetisations}b). The obtained CEF parameters (Table.~\ref{tab:table1}) obey a general rule, that a ratio of the rhombic anisotropy $E$ = $B_2^2$ to the axial anisotropy $D$ = 3$\times$$B_{\rm 2}^{\rm 0}$, $E/D$ is always positive and acquires values between 0 and 1/3. CEF values from the Table.~\ref{tab:table1} provide $E/D$ $\sim$ 0.2. Corresponding energy scheme (Fig.~\ref{Fig:Models}b) is characterized by rather large separation between the third and fourth energy doublet which should manifest in the specific heat as a separate maximum (Fig.~\ref{Fig:HC}a). The absence of such maximum in the experiment cannot be explained by the influence of magnetic correlations which become important below 1 K. Apparently, the obtained CEF parameters can be treated as effective, since the analysis did not involve other subtle effects as the aforementioned lower local symmetry, demagnetization effects and/or potential misalignment of the sample by a few degrees from the crystallographic axes. In accord with previous susceptibility studies ~\cite{Feher1988}, the effective CEF parameters reflect the predominance of the uniaxial anisotropy providing nearly identical magnetization values for hard (crystallographic axis $b$) and medium (crystallographic axis $c$) axes. In comparison with the previous estimate, $B_{\rm 2}^{\rm 0}$ = -0.089 K ~\cite{Feher1988}, the lower value of the uniaxial term ($B_{\rm 2}^{\rm 0}$ = - 0.0661 K, Table.~\ref{tab:table1}) indicates weaker uniaxial anisotropy in CsGd(MoO$_4$)$_2$. 

\begin{table}
\caption{\label{tab:table1}Effective CEF parameters obtained from the analysis of magnetization curves (in units of Kelvin).}

\begin{tabular}{ccccc}
\\ \hline \rule{0pt}{9pt}
$B$$_2^0$ & $B$$_2^2$ & $B$$_4^0$ & $B$$_4^2$ & $B$$_4^4$\\ 
-0.0661 &-0.0398 &-0.0013 &-0.0033 & -0.0116 
\\ \hline \rule{0pt}{9pt}
$B$$_6^0$ & $B$$_6^2$ & $B$$_6^4$ & $B$$_6^6$ & \\
1.0918$\times$10$^{-4}$ &-0.0018 &0.0017 &7.6261$\times$10$^{-4}$ &   
\\ \hline \rule{0pt}{9pt}
\end{tabular}
\end{table}

The actual significance of low-symmetry CEF components was verified by the simultaneous fitting of magnetization curves at $T$ = 2 K in both orientations with a CEF model considering only $B_{\rm 2}^{\rm 0}$ parameter (Fig.~\ref{Fig:Magnetisations}a). The procedure yielded $B_{\rm 2}^{\rm 0}$ = -0.0557 K suggesting even weaker CEF splitting (Fig.~\ref{Fig:Models}b). The agreement with magnetization curves does not achieve such quality as the description within the D$_2$ symmetry, but it can be improved by considering demagnetization effects with the average demagnetizing factor $N$ $\sim$ 0.2. Apparently, the simplified model provides much better description of the specific heat than the effective model with D$_2$ symmetry (Fig.~\ref{Fig:HC}a). Possibly, the weak low-symmetry components can partially interfere with magnetic correlations of comparable strength. 

Moreover, the interplay of lattice vibrations and the CEF electronic states with the energies corresponding to the model with $B_{\rm 2}^{\rm 0}$ = -0.0557 K can elucidate anomalous behavior of the phonon mean free path, $L$, below 1 K (Fig.~\ref{Fig:HC}b). Unlike previous analysis ~\cite{Tkac2013}, the existence of plateau below 1 K should result from the scattering of phonons on the internal surfaces of this highly anisotropic material with a layered structure. While the low-temperature end of the plateau nicely correlates with the magnetic phase transition at 0.45 K, further decrease of $L$ with a minimum appearing at 0.2 K can be attributed to the one-magnon-one-phonon resonance phonon scattering \cite{berman1976thermal}. Within the Debye model, the acoustic phonons with the energies $\hbar \omega$ $\sim$ 3.8$k_{\rm B}$$T$ provide the greatest contribution to the phonon heat capacity. Accordingly, the phonons with the CEF energies dominate the phonon spectrum at temperatures as denoted in Fig.~\ref{Fig:HC}b. The correlation between the hypothetic temperatures and the observed anomalies is apparent.

\subsection{Magnetocaloric effect}

\begin{figure}
\includegraphics{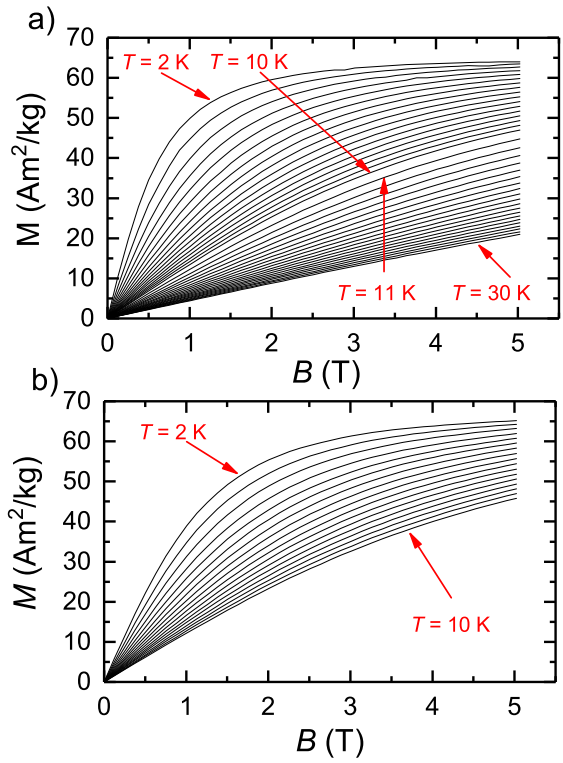} \\
\caption{\label{Fig:MagnetisationsAll} (Color online) Isothermal magnetization curves of CsGd(MoO$_4$)$_2$ for (a) $B$$\parallel$$a$, measured with the temperature step 0.5 K and 1 K for the temperature interval 2-10 K and 11-30 K, respectively (b) $B$$\parallel$$c$, measured with the temperature step 0.5 K for the temperature interval 2-10 K. }
\end{figure}

Isothermal magnetization curves in the magnetic field oriented along the $a$ and $c$ axes are depicted in Fig.~\ref{Fig:MagnetisationsAll}.
The data were used for the calculation of the isothermal magnetic entropy change, $\Delta S_{\rm M}$, applying the Maxwell relation \cite{Gschneidner2005RPP}
\begin{eqnarray}
\Delta S_M(T, \Delta B) = \int^{B_f}_{B_i} \frac{\partial M(T,B)}{\partial T}dB,
\label{eq:Magnetic_entropy}
\end{eqnarray} 
where $B_{\rm i}$ and $B_{\rm f}$ represent initial and final magnetic field, respectively. Temperature dependence of -$\Delta S_{\rm M}$ derived from the experimental magnetization data for $B_{\rm i}$ = 0 and several values of $B_{\rm f}$ applied along the $a$ and $c$ axes is shown in Fig.~\ref{Fig:Isothermal entropy change}. 

\begin{figure}
\includegraphics{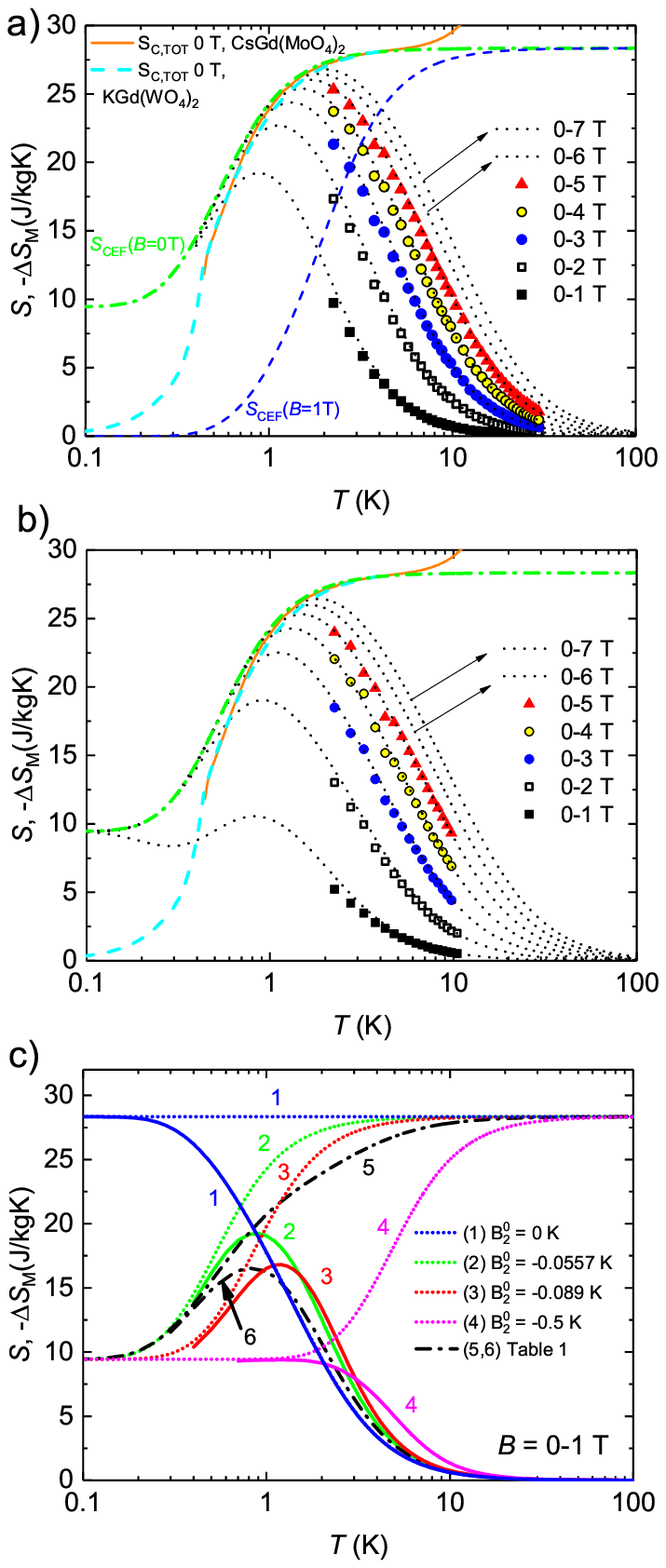} \\
\end{figure}

\begin{figure}
\caption{\label{Fig:Isothermal entropy change} (Color online) (a) Temperature dependence of the entropy and isothermal entropy change in CsGd(MoO$_4$)$_2$ in different magnetic fields, $B$$\parallel$$a$. Symbols represent $-\Delta S_{\rm M}$ values obtained from experimental magnetization curves, dotted lines represent $-\Delta S_{\rm M}$ values calculated from the CEF energies corresponding to $B_{\rm 2}^{\rm 0}$ = -0.0557 K. A solid and long-dashed line represent a total entropy of CsGd(MoO$_4$)$_2$ and KGd(WO$_4$)$_2$,  respectively.  For more details, see text.  Dashed-dotted and short-dashed line represent magnetic entropy in zero field and 1 T, respectively, calculated from the CEF energies corresponding to $B_{\rm 2}^{\rm 0}$ = -0.0557 K. (b) Temperature dependence of the entropy and isothermal entropy change in CsGd(MoO$_4$)$_2$ in different magnetic fields, $B$$\parallel$$c$. Symbols and lines have the same meaning as in (a). (c) Temperature dependence of the entropy in $B$ = 0 T (dotted lines) and isothermal entropy change in the magnetic field ($B$$\parallel$$a$) changing from zero to 1 T (solid lines) calculated for the different strength of $B_{\rm 2}^{\rm 0}$. Dash-dotted lines (5) and (6) represent $S$ and $-\Delta S_{\rm M}$, respectively, corresponding to the parameters in Table.~\ref{tab:table1}.}
\end{figure}

It is obvious, that for all magnetic field changes, the entropy change achieves maximum out of the experimental temperature window. To complete the information, theoretical -$\Delta S_{\rm M}$ values were calculated from the CEF model with uniaxial symmetry for $B_{\rm 2}^{\rm 0}$ = -0.0557 K in a wide temperature range for the same magnetic fields (Fig.~\ref{Fig:Isothermal entropy change}). Excellent agreement with experimental data supports the prevalence of the CEF with uniaxial symmetry affecting magnetocaloric properties at temperatures far above $T_{\rm C}$. Theoretical -$\Delta S_{\rm M}$ curves achieve maximum below 2 K, shifting towards low temperatures for decreasing magnetic field $B_{\rm f}$. Closer inspection revealed that at low temperatures, all theoretical -$\Delta S_{\rm M}$ curves merge into a universal curve with a nonzero value. This rather artificial behavior results from a two-fold degeneracy of the CEF ground state in zero magnetic field. As can be seen, the universal curve enveloping the low-temperature parts of the theoretical -$\Delta S_{\rm M}$ curves represents a theoretical entropy in zero magnetic field, $S_{\rm CEF}$, derived from the CEF model with the aforementioned uniaxial symmetry (Fig.~\ref{Fig:Isothermal entropy change}). At high temperatures, the theoretical entropy (in units of J/kg K) saturates to a maximum value $S_{\rm max}$ = $ln(2S+1) R/M$ ($R$ and $M$ represent a gas constant and a molar mass, respectively). For CsGd(MoO$_4$)$_2$ parameters, $S_{\rm max}$ achieves 28.34 J/kg K. At low temperatures $S_{\rm CEF}$ attains minimum value, ln(2)$R/M$ = 9.45 J/kg K.  

Concerning the calculation of the experimental entropy in zero magnetic field, $S_{\rm c,tot}$($B$ = 0 T), total specific heat data were used in a wide temperature region. Since the lowest temperature achieved in the low-temperature experiment \cite{Feher1988} was 0.42 K i.e., 0.95 $T_{\rm C}$, unknown amount of magnetic entropy still remained below a phase transition. To avoid large uncertainty introduced by any artificial extrapolation of the specific heat down to zero temperature, the experimental data of a similar KGd(WO$_4$)$_2$ compound \cite{Borowiec2013} were used for the calculation of the missing entropy in the whole ordered region. The excellent agreement of both specific heat datasets suggests close similarity of dipolar interactions and CEF symmetry in both compounds (inset of Fig.~\ref{Fig:Magnetisations}a). 

The comparison of the experimental entropy (comprising lattice and magnetic contribution) and $S_{\rm CEF}$ in zero magnetic field (Fig.~\ref{Fig:Isothermal entropy change}) point at the prevalence of CEF contribution above 2 K. A deviation of the experimental entropy from $S_{\rm CEF}$ gradually developing below 1 K indicates the onset of magnetic correlations resulting in a phase transition accompanied with a sudden drop of the experimental entropy at $T_{\rm C}$.   
Concerning the $S_{\rm CEF}$ in nonzero magnetic field, the application of external magnetic field removes the two-fold CEF ground-state degeneracy. Corresponding energy gap projects to the exponential decrease of $S_{\rm CEF}$ at lowest temperatures. As a consequence, $S_{\rm CEF}$ calculated for $B$ = 1 T achieves nearly zero values at temperatures below 0.5 K (Fig.~\ref{Fig:Isothermal entropy change}a). Increasing magnetic field enforces the exponential decrease, shifting the region of nearly zero $S_{\rm CEF}(B)$ values towards higher temperatures. This result suggests, that alike theoretical -$\Delta S_{\rm M}$ curves merge at low temperatures into a universal curve $S_{\rm CEF}$($B$ = 0 T), all real -$\Delta S_{\rm M}$ vs. $T$ dependencies are expected to merge into a universal curve identical with the experimental entropy in zero magnetic field. As was shown in ref. \cite{Tkac2015}, the experimental entropy forms an envelope of a low-temperature side of the experimental -$\Delta S_{\rm M}$ maxima. At higher temperatures, at least above 2 K, the contribution of crystal field prevails in all studied magnetic fields, leading to large conventional magnetocaloric effect. In the field 5 T applied along the easy axis, $\Delta S_{\rm max}$ = 26 J/kg K already achieves nearly 92 $\%$ of $S_{\rm max}$. 

Concerning the isothermal entropy change in $B$$\parallel$$c$, as expected from magnetization data (Fig.~\ref{Fig:Magnetisations}), the values are comparable with those in the field applied along the easy axis (Fig.~\ref{Fig:Isothermal entropy change}b).

Noticeable differences between -$\Delta S_{\rm M}$ in $B$$\parallel$$a$ and $B$$\parallel$$c$ appearing in magnetic fields lower than 2 T can be associated with the CEF effects. A closer examination of the CEF impact on the magnetic entropy in zero magnetic field as well as -$\Delta S_{\rm M}$ parameters in the relatively low magnetic field 1 T applied along the $a$ axis, revealed rather high sensitivity of MCE parameters (Fig.~\ref{Fig:Isothermal entropy change}c). Neglecting any magnetic correlations, the absence of CEF effect ($B^0_2$ = 0 K) leads to the full spin degeneracy in zero magnetic field while CEF with any symmetry partially removes the spin degeneracy in Gd$^{3+}$ ion. Assuming uniaxial CEF symmetry, the strengthening of magnetic anisotropy (the increase of $B^0_2$ parameter) leads to lowering of maximal  -$\Delta S_{\rm M}$ values. What is more, the maxima shift towards higher temperatures. Analogical effect can be expected for CEF with lower symmetry. As can be seen, despite the nearly isotropic nature of Gd$^{3+}$ ion, CEF plays important role in MCE parameters at temperatures comparable with the zero-field splitting (Fig.~\ref{Fig:Models}b).   

\begin{figure}
\includegraphics{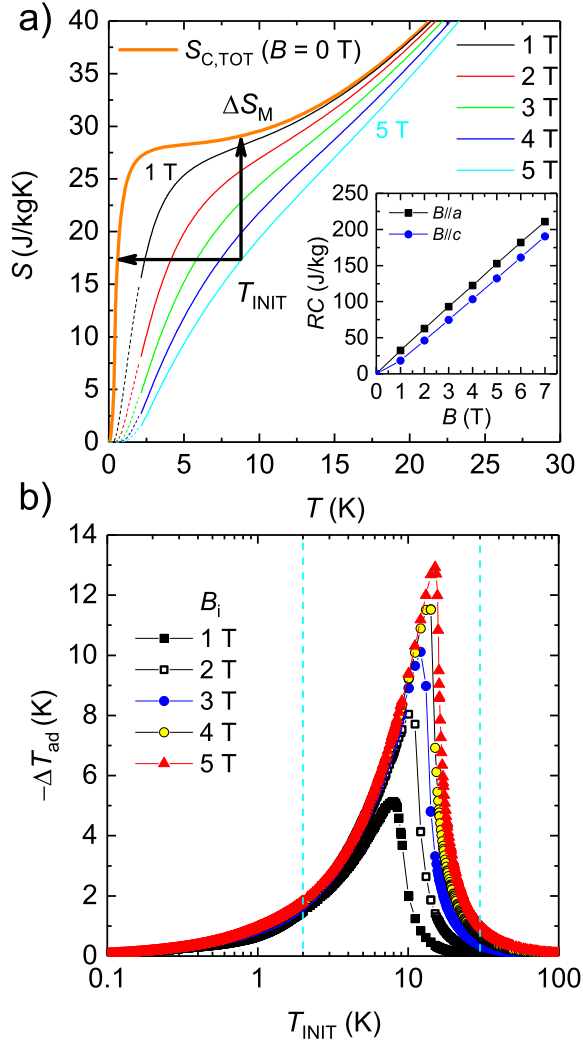} \\
\caption{\label{Fig:Enthropy_Ba} (a) Temperature dependence of the total entropy of CsGd(MoO$_4$)$_2$ comprising magnetic and lattice contribution. A thick solid line represents the experimental entropy in zero magnetic field. Thin solid lines represent experimental entropies in magnetic field applied along the $a$ axis. The dashed lines represent estimates using CEF theory (for more details, see text). The horizontal arrow shows how much the sample is cooled down from the initial temperature, $T_{\rm INIT}$, in the isentropic process, while the vertical arrow demonstrates an example of the isothermal entropy change. Inset: Field dependence of the refrigerant capacity $RC$. (b) Adiabatic temperature change as a function of the initial temperature during the isentropic change from nonzero initial magnetic field, $B_{\rm i}$, to zero $B_{\rm f}$. Vertical dashed lines denote low-temperature and high-temperature interval in which the estimations of $\Delta T_{\rm ad}$ were calculated from CEF values (see text).}
\end{figure}

The refrigerant capacity, $RC$, was calculated from the experimental -$\Delta S_{\rm max}$ values \cite{Gschneidner2005RPP}, using a relation $RC$ = $\int^{T_{hot}}_{T_{cold}}$ $\left| \Delta S_{\rm M}(T) \right| dT$, where $T_{\rm cold}$ and $T_{\rm hot}$ a working temperature interval of the refrigerant (Fig.~\ref{Fig:Enthropy_Ba}a inset). We used $T_{cold}$ = 0.4 K, while $T_{\rm hot}$ is a temperature, at which the quantity -$\Delta S_{\rm M}$ reaches half of the maximum value. 

In the field applied along the $a$ (easy) axis, $RC$ achieves 215 J/kg for 7 T, whereas for the same field applied along the $c$ axis, $RC$ $\sim$ 200 J/kg.

The relative independence of MCE on the orientation of magnetic field ($B$ $\gtrsim $ 1 T) allows the use of CsGd(MoO$_4$)$_2$ for practical applications in the form of powder. Unlike single crystals, this rather comfortable form of refrigerant is not so sensitive to introducing strains or other imperfections.

The adiabatic temperature change, $- \Delta T_{\rm ad}$ is associated with the isentropic change from nonzero initial magnetic field, $B_{\rm i}$, to the final zero value and represents another important parameter of MCE. The procedure of the determination of $-\Delta T_{\rm ad}$ is depicted in Fig.~\ref{Fig:Enthropy_Ba}a. Experimental entropy curves in nonzero magnetic field, $S(B,T)$, were constructed using the relation \cite{Tkac2015} $S(B,T)$ = $S_{\rm C,TOT}$$(B = 0,T)$-$\left| \Delta S_{\rm M}(B,T) \right|$ at temperatures down to 2 K. At lower temperatures, unknown experimental values $\left| \Delta S_{\rm M}(B,T) \right|$ were approximated by $\Delta S_{\rm CEF}$ values. Corresponding estimations of $- \Delta T_{\rm ad}$ values achieve maximum at temperatures around 10 K and the maximum shifts towards higher temperatures for higher initial magnetic field (Fig.~\ref{Fig:Enthropy_Ba}b). As can be seen, setting the sample to the initial temperature $T_{\rm INIT}$ = 15 K and field 5 T, the material is cooled down to 2 K during adiabatic change of magnetic field to zero value. 

Despite good MCE parameters achieved in CsGd(MoO$_4$)$_2$, there still exists a possibility how to improve magnetocaloric properties of the double molybdates. Concerning the paramagnetic phase, large magnetocaloric effect given by -$\Delta S_{\rm M}$ values, depends on the interplay of a few mechanisms. The high density of magnetic moments (small molar mass $M$) and high spin values enhance maximum magnetic entropy $S_{\rm max}$ = $ln(2S+1) R/M$  which represents the upper limit for -$\Delta S_{\rm M}$. However, increasing density of the magnetic moments leads to the enhancement of magnetic interactions between magnetic moments which lower magnetic entropy as well as -$\Delta S_{\rm M}$ at temperatures where the magnetic correlations dominate. As was already demonstrated in Fig.~\ref{Fig:Isothermal entropy change}c, crystal electric field has similar effect.

Examples of magnetocaloric properties of selected Gd-based oxides with various $M$ parameters and the strength of magnetic correlations are given in the Table.~\ref{tab:table list}. As can be seen, for some materials a large difference between $S_{\rm max}$ and maximum -$\Delta S_{\rm M}$ is observed as a result of the aforementioned effects. In the majority of the systems (Table.~\ref{tab:table list}) the maximal value of -$\Delta S_{\rm M}$ is achieved at temperatures $T_{\rm max}$ $\lesssim$ $T_{\rm C}$ and it does not reach the value of $S_{\rm max}$. The significant reduction of -$\Delta S_{\rm M}$ can be ascribed to the influence of  strong magnetic correlations with corresponding large value of $T_{\rm C}$. Apparently, only few systems can use full potential of the magnetic degrees of freedom.

Thus, considering double molybdates, already a simple replacement of the Cs$^+$ ion by other alkali-metal ions M$^+$ with smaller mass, as Li, Na, K or Rb, should provide for corresponding MGd(MoO$_4$)$_2$ compounds maximal magnetic entropy $S_{\rm max}$ = 35.7, 34.6, 33.5 and 30.7 J/kgK, respectively. As can be seen from the Table.~\ref{tab:table list}, further increase of -$\Delta S_{\rm M}$ would require hypothetical compounds with higher number of Gd ions per formula and at the same time weak magnetic correlations between Gd ions.

\begin{table}
\caption{\label{tab:table list} Magnetocaloric properties of selected Gd-based oxides. $S_{\rm max}$ is calculated for $S$ = 7/2.}

\begin{tabular}{cccccc}
\\ \hline \rule{0pt}{9pt}
Compound & -$\Delta S_{\rm M}$ & $T_{\rm max} $ & $B_{\rm f} $ & $T_{\rm C}$  & $S_{\rm max}$ = ln(2$S$+1)$R$/$M$ \\
& (J/kgK) & (K) & (T) & (K) & (J/kgK)
\\ \hline \rule{0pt}{9pt}
GdMnO$_3$ \cite{Aditya2015} & 31 & 7 & 8 & 42, 23, 5.2 & 66.45\\
Gd$_2$O$_3$ \cite{PAUL2016182}& 10.7 & 2$^*$ & 5 & 3.9 & 95.4\\
Gd$_3$Fe$_5$O$_{12}$ \cite{Phan2009}& 2.45 & 35 & 3 & 90 & 55\\
Gd$_3$Ga$_5$O$_{12}$ \cite{Hamilton2014}& 24 & 2$^*$ & 5 & & 51.24\\
Gd$_3$Al$_5$O$_{12}$ \cite{Hamilton2014}& 28 & 2$^*$ & 5 & & 64.95\\
GdAlO$_{3}$ \cite{Mahana2017}& 40.9 & 2 & 9 & 40, 25, 3.9 & 74.75\\
GdVO$_{4}$ \cite{Dey2017}& 41.1 & 3 & 5 & 2.4 & 63.52 \\
GdCrO$_{4}$ \cite{Midya2014}& 28 & 20 & 7 & 20 & 63.28\\
GdPO$_{4}$ \cite{Palacios2014}& 62 & 2.1 & 7 & 0.77 & 68.55
\\ \hline
\end{tabular}
$^*$ values obtained at lowest experimental temperature.
\end{table}

\section{Conclusion} 

In conclusion, we studied magnetic and magnetocaloric properties of the single crystal of CsGd(MoO$_4$)$_2$. The analysis of specific heat and magnetization provided refinement of CEF parameters indicating the dominance of uniaxial symmetry of local crystal field. Maximum values of the isothermal entropy change in magnetic fields up to 5 T are expected to occur at temperatures around 2 K. It should be noted that -$\Delta S_{\rm M}$  achieves 18 J/kgK  already for the field 1 T, while for 7 T, maximal -$\Delta S_{\rm M}$ = 26.8 J/kgK with a refrigerant capacity of 215 J/kg. The absence of thermal hysteresis and the losses due to eddy currents as well as good chemical stability and high thermal conductivity makes the compound CsGd(MoO$_4$)$_2$ attractive for magnetic refrigeration at low temperatures.

Last but not least, our simulations of crystal field effect showed that stronger CEF lowers the maximal value of -$\Delta S_{\rm M}$, and shifts the position of the -$\Delta S_{\rm M}$ maximum towards higher temperatures. Thus, the strength of CEF can control the size of MCE as well as a working temperature interval.\\

\section{Acknowledgments}
This work has been supported by VEGA grant 1/0269/17, projects APVV-0132-11, 14-0073 and ERDF EU project No. ITMS26220120047. Financial support of US Steel DZ Energetika is greatly acknowledged.



\begin{thebibliography}{10}

\bibitem{Tishin2003}
A.~M. Tishin,
\newblock {\em The Magnetocaloric Effect and its Applications} (CRC Press,
  Abingdon, 2003).

\bibitem{Tegus2002}
O.~Tegus, E.~Bruck, K.~H.~J. Buschow, and F.~R. de~Boer,
\newblock Nature {\bf 415}, 150 (2002).

\bibitem{Pecharsky1997}
V.~K. Pecharsky and K.~A. Gschneidner, Jr.,
\newblock Phys. Rev. Lett. {\bf 78}, 4494 (1997).

\bibitem{Pecharsky2002}
V.~K. Pecharsky, A.~Pecharsky, and K.~G. Jr.,
\newblock Journal of Alloys and Compounds {\bf 344}, 362  (2002),
\newblock Proceedings of the Rare Earths` 2001 Conference.

\bibitem{Wu2015}
Y.~Wu, J.~Wang, H.~Hua, C.~Jiang, and H.~Xu,
\newblock Journal of Alloys and Compounds {\bf 632}, 681  (2015).

\bibitem{Aksoy2008}
S.~Aksoy {\em et~al.},
\newblock Journal of Alloys and Compounds {\bf 460}, 94  (2008).

\bibitem{Jayaraman2011}
T.~V. Jayaraman, L.~Boone, and J.~E. Shield,
\newblock Journal of Alloys and Compounds {\bf 509}, 1411  (2011).

\bibitem{Min2014}
J.~Min, X.~Zhong, Z.~Liu, Z.~Zheng, and D.~Zeng,
\newblock Journal of Alloys and Compounds {\bf 606}, 50  (2014).

\bibitem{Wang2016}
Z.~Wang, P.~Yu, Y.~Cui, and L.~Xia,
\newblock Journal of Alloys and Compounds {\bf 658}, 598  (2016).

\bibitem{Sedlakova2009}
L.~Sedl{\'{a}}kov{\'{a}} {\em et~al.},
\newblock Journal of Alloys and Compounds {\bf 487}, 425  (2009).

\bibitem{Chen2013}
Y.-C. Chen {\em et~al.},
\newblock Chemistry – A European Journal {\bf 19}, 13504 (2013).

\bibitem{Chen2014}
Y.-C. Chen {\em et~al.},
\newblock J. Mater. Chem. A {\bf 2}, 9851 (2014).

\bibitem{Guo2013}
F.-S. Guo {\em et~al.},
\newblock Chemistry – A European Journal {\bf 19}, 14876 (2013).

\bibitem{Gschneidner2005}
K.~A. GschneidnerJr, V.~K. Pecharsky, and A.~O. Tsokol,
\newblock Reports on Progress in Physics {\bf 68}, 1479 (2005).

\bibitem{Zelenakova2016}
A.~Zele{\v{n}}{\'{a}}kov{\'{a}}, P.~Hrubov{\v{c}}{\'{a}}k, O.~Kapusta,
  V.~Zele{\v{n}}{\'{a}}k, and V.~Franco,
\newblock Applied Physics Letters {\bf 109}, 122412 (2016),
  http://dx.doi.org/10.1063/1.4963267.

\bibitem{Evangelisti2009}
M.~Evangelisti {\em et~al.},
\newblock Phys. Rev. B {\bf 79}, 104414 (2009).

\bibitem{bartolome2013molecular}
S.~Bartolome, F.~Luis, and J.~Fern{\'a}ndez,
\newblock {\em {M}olecular {M}agnets: {P}hysics and {A}pplications}NanoScience
  and Technology (Springer Berlin Heidelberg, 2013).

\bibitem{Wang2014}
F.~Wang, F.~ying Yuan, J.~zhi Wang, T.~fu~Feng, and G.~qi~Hu,
\newblock Journal of Alloys and Compounds {\bf 592}, 63  (2014).

\bibitem{Sibille2014}
R.~Sibille, E.~Didelot, T.~Mazet, B.~Malaman, and M.~François,
\newblock APL Materials {\bf 2}, 124402 (2014),
  http://dx.doi.org/10.1063/1.4900884.

\bibitem{Palacios2014}
E.~Palacios {\em et~al.},
\newblock Phys. Rev. B {\bf 90}, 214423 (2014).

\bibitem{Shi2014}
P.~Shi, Z.~Xia, M.~S. Molokeev, and V.~V. Atuchin,
\newblock Dalton Trans. {\bf 43}, 9669 (2014).

\bibitem{Zhao2015}
W.~Zhao {\em et~al.},
\newblock RSC Adv. {\bf 5}, 34730 (2015).

\bibitem{Devakumar2017}
B.~Devakumar, P.~Halappa, and C.~Shivakumara,
\newblock Dyes and Pigments {\bf 137}, 244  (2017).

\bibitem{Tkac2013}
V.~Tk{\'{a}}{\v{c}} {\em et~al.},
\newblock Journal of Physics: Condensed Matter {\bf 25}, 506001 (2013).

\bibitem{Tkac2014}
V.~Tk{\'{a}}{\v{c}} {\em et~al.},
\newblock Journal of Alloys and Compounds {\bf 591}, 100  (2014).

\bibitem{Stefanyi1988}
P.~Stef{\'{a}}nyi, A.~Feher, A.~Orend{\'{a}}{\v{c}}ov{\'{a}}, E.~Anders, and
  A.~Zvyagin,
\newblock Journal of Magnetism and Magnetic Materials {\bf 73}, 129  (1988).

\bibitem{Feher1988}
A.~Feher {\em et~al.},
\newblock Sov. J. Low. Temp. Phys. {\bf 14}, 723 (1988).

\bibitem{Zaboj1992}
R.~Z{\'{a}}boj, P.~Stef{\'{a}}nyi, and A.~Feher,
\newblock Journal of Magnetism and Magnetic Materials {\bf 104}, 953  (1992).

\bibitem{Anders1995}
A.~G. Anders, S.~V. Volotskii, S.~V. Startsev, A.~Feher, and
  A.~Orend{\'{a}}{\v{c}}ov{\'{a}},
\newblock Low. Temp. Phys. {\bf 21}, 38 (1995).

\bibitem{Tibenska2010}
K.~Tibensk{\'{a}} {\em et~al.},
\newblock Acta Physica Polonica A. {\bf 118}, 971 (2010).

\bibitem{Vinokurov1972}
V.~Vinokurov and P.~Klevtsov,
\newblock Sov. Phys. Crystallogr. (Engl. Transl.) {\bf 17}, 127 (1972).

\bibitem{database}
Y.~Xu, M.~Yamazaki, and P.~Villars,
\newblock Japanese Journal of Applied Physics {\bf 50}, 11RH02 (2011).

\bibitem{Abragam1970}
A.~Abragam and B.~Bleaney,
\newblock {\em Electron Paramagnetic Resonance of Transition Ions} (Clarendon
  Press, 1970).

\bibitem{Borowiec2013}
M.~T. Borowiec {\em et~al.},
\newblock Central European Journal of Physics {\bf 11}, 394 (2013).

\bibitem{berman1976thermal}
R.~Berman,
\newblock {\em Thermal Conduction in Solids} (Clarendon Press, 1976).

\bibitem{Gschneidner2005RPP}
K.~A. GschneidnerJr, V.~K. Pecharsky, and A.~O. Tsokol,
\newblock Reports on Progress in Physics {\bf 68}, 1479 (2005).

\bibitem{Tkac2015}
V.~Tk\'a\ifmmode~\check{c}\else \v{c}\fi{} {\em et~al.},
\newblock Phys. Rev. B {\bf 92}, 024406 (2015).

\bibitem{Aditya2015}
A.~A. Wagh, K.~G. Suresh, P.~S.~A. Kumar, and S.~Elizabeth,
\newblock Journal of Physics D: Applied Physics {\bf 48}, 135001 (2015).

\bibitem{PAUL2016182}
R.~Paul {\em et~al.},
\newblock Journal of Magnetism and Magnetic Materials {\bf 417}, 182  (2016).

\bibitem{Phan2009}
M.~H. Phan {\em et~al.},
\newblock Journal of Physics D: Applied Physics {\bf 42}, 115007 (2009).

\bibitem{Hamilton2014}
A.~C.~S. Hamilton, G.~I. Lampronti, S.~E. Rowley, and S.~E. Dutton,
\newblock Journal of Physics: Condensed Matter {\bf 26}, 116001 (2014).

\bibitem{Mahana2017}
S.~Mahana, U.~Manju, and D.~Topwal,
\newblock Journal of Physics D: Applied Physics {\bf 50}, 035002 (2017).

\bibitem{Dey2017}
K.~Dey, A.~Indra, S.~Majumdar, and S.~Giri,
\newblock J. Mater. Chem. C {\bf 5}, 1646 (2017).

\bibitem{Midya2014}
A.~Midya, N.~Khan, D.~Bhoi, and P.~Mandal,
\newblock Journal of Applied Physics {\bf 115}, 17E114 (2014),
  http://dx.doi.org/10.1063/1.4861680.

\end{thebibliography}


\end{document}